\documentclass[a4paper,oneside]{saip}
\usepackage{helvet}          
\usepackage{times}           
\usepackage{courierten}   
\usepackage[T1]{fontenc} 

\usepackage{graphicx}  
\usepackage{url}
\usepackage{geometry} 
\usepackage[labelformat=simple]{caption}
\usepackage{subcaption}
\usepackage[table,x11names]{xcolor}
\usepackage{amsmath}
\usepackage{siunitx}
\usepackage{academicons}
\usepackage{fontawesome5}
\usepackage[section]{placeins}
\usepackage{pdfpages}
\usepackage{appendix}
\definecolor{orcidlogocol}{HTML}{A6CE39}

\begin{document}

\title{\textbf{Developing a Sky-End Fibre Aperture for the AMASE Fibre Bundle}}

\author{Goratamang Gaedie \href{https://orcid.org/0009-0009-0634-6073}{\textcolor{orcidlogocol}{\faOrcid}} $^{1,2}$, Sabyasachi Chattophadyay$^{1,2}$, and Nidhi Mehandirrata$^{2,3}$}

\affil{$^1$Centre for Space Research, North-West University, Potchefstroom 2520, South Africa}
\affil{$^2$South African Astronomical Observatory, Observatory 7925, Cape Town, South Africa}
\affil{$^3$Department of Astronomy, University of Cape Town, 7700, South Africa}

\email{\url{ggaedie@gmail.com}}


\begin{abstract}
The Affordable Multiple-Aperture Spectroscopy Explorer (AMASE) project aims to enable high-resolution spectroscopy of ionised gas in galaxies. However, fabricating ultradense fibre bundles with micrometre-level precision presents significant challenges. We detail a fabrication process that combines precision machining, stress-reducing handling, and iterative polishing. Although the final prototype did not meet the optical performance targets, our methods reduced manual labour, improved reproducibility, and identified critical failure modes, providing essential information for future astrophotonic instrument designs.
\end{abstract}

\section{Introduction}\label{sec:intro}
Over the past decade, there has been a growing emphasis on understanding the physical processes governing the interstellar medium (ISM) and star formation in the Milky Way and nearby galaxies \citep{Kennicutt2012,Ferriere2001}. Integral Field Spectroscopy (IFS) has revolutionised this effort, enabling the study of spatially resolved kinematics, ionisation states, and chemical abundances across diverse galactic environments. Studies like MaNGA (Mapping Nearby Galaxies at Apache Point Observatory, which is part of SDSS-IV and conducts integral-field spectroscopy on approximately 10,000 nearby galaxies) \citep{Bundy2015} and SAMI (Sydney-AAO Multi-object Integral-field spectrograph, observing thousands of galaxies in various environments using multi-object IFU spectroscopy) \cite{Croom2012, Bryant2015} have offered vital information on kiloparsec scales. However, these surveys lack spatial and spectral resolution to probe the fine structure and turbulence within individual H\textsc{II} regions at sub-parsec levels, which is critical to understanding star formation feedback \cite{Zhou2017,Oh2022}.\\

The Affordable Multiple Aperture Spectroscopy Explorer (AMASE) project addresses this gap through a novel, cost-effective design that combines an array of multifibre spectrographs with telephoto lenses \citep{yan2020}. AMASE targets a spectral resolution of $R \sim 15{,}000$ and a spatial resolution ($\rm \sigma \sim 8.5$ kms$^{-1}$) capable of dissecting ionised gas structures from 0.1 to 100 parsecs in the Local Group galaxies. Key science goals of AMASE include: (1) characterising feedback-driven outflows in star-forming regions \citep{Law2022,Zhou2017}; (2) disentangling ionisation sources in diffuse ionised gas \citep{Zhou2017}; (3) calibrating strong-line metallicity diagnostics; and (4) improving photoionisation models with spatially resolved emission line data \citep{Oh2022}. Achieving these goals requires precise control over the optical alignment and integrity of the fibre bundles that couple the spectrographs to the sky.\\

We present the design, fabrication, and metrological validation of the sky-end fibre instrument cable, a key component of AMASE. This system enables high-resolution, integral-field spectroscopy of ionised gas structures in the Milky Way and Local Group galaxies. A high density configuration of 547 fibres within a 1.858 mm regular hexagonal aperture achieves a core area-to-hexagon area ratio of approximately 50\%, maximising photon throughput. We report on the design, metrology, and analysis of the quality of the fibre bundle.\\


\section{Fabrication Methodology}\label{sec:method}

The AMASE instrument uses 547 fibres with a core diameter of \SI{50}{\micro\metre}, each providing a spatial resolution of approximately 25.24 arcseconds on-sky. These fibres cover a hexagonal on-sky footprint with a long-axis span of 19.3 arcminutes, equivalent to a total field of view of roughly 79 arcmin\textsuperscript{2} per spectrograph. The fibres are arranged in a hexagonal close-packed (HCP) geometry to maximise the fill factor and minimise interstitial space. Figure \ref{AppendixA} is an illustration of the design of the hexagonal apertures used for AMASE.\\

\begin{figure}[h!]
    \centering
    \includegraphics[width=0.5\linewidth]{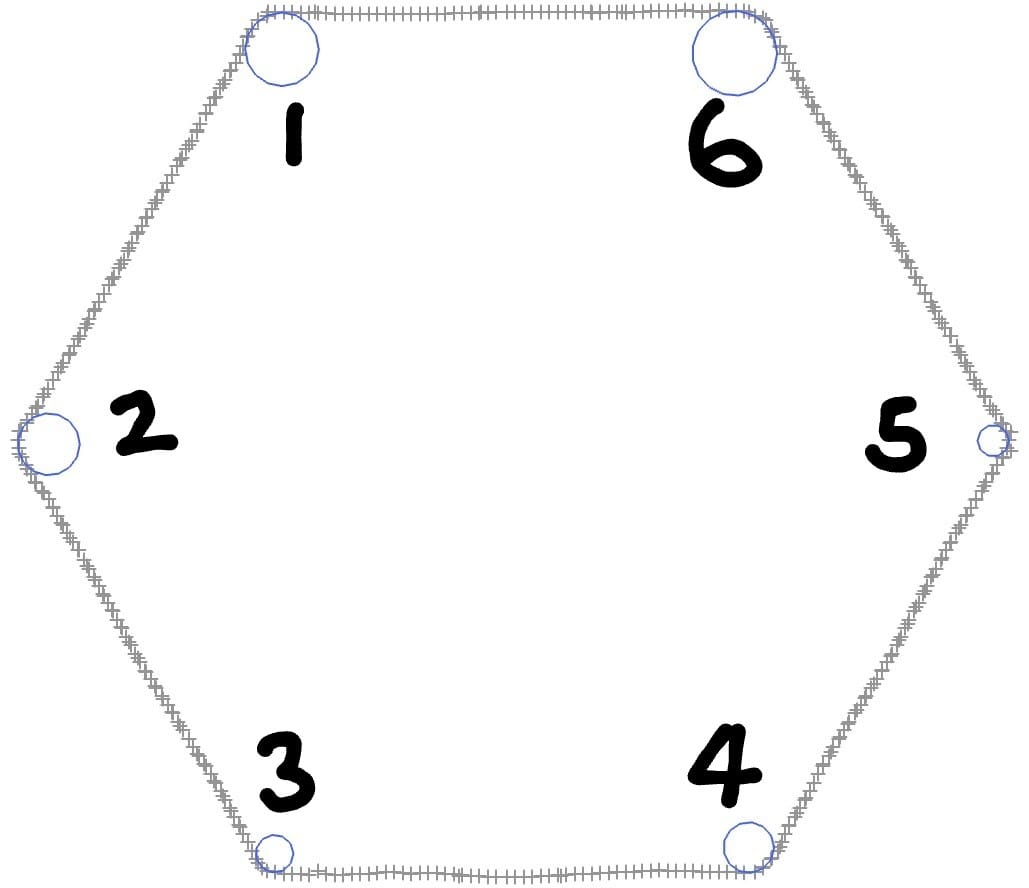}
    \caption{Microscopic (shadowgraph) inspection of the aperture quality and dimensional conformity. At each corner of the hexagonal aperture, the smallest possible circle has been fitted. The radii of these circles are: (1) 0.081 mm, (2) 0.068 mm, (3) 0.042 mm, (4) 0.056 mm, (5) 0.035 mm, and (6) 0.094 mm.}
    \label{fig:circlefit}
\end{figure}

The fibre holding aperture has a nominal diagonal dimension of \SI{1.894}{\milli\metre}. In Figure~\ref{fig:circlefit}, this corresponds to the diagonal distance measured between the corner at circle one and the opposite corner at circle four. However, practical assembly constraints necessitated additional clearance to account for fibre-dimensional variability. The clearance was defined as follows:

\begin{equation}
\text{Clearance} = \Delta d \times \sqrt{N}
\end{equation}

where $\Delta d$ is the tolerance for the diameter of the fibre and $N$ is the number of fibres along the dimension. Given (from the fibre supplier) a typical fibre diameter tolerance of $\pm\SI{1}{\micro\metre}$, the required clearance becomes:

\begin{equation}
\text{Clearance} = \SI{1}{\micro\metre} \times \sqrt{27} \approx \SI{5.20}{\micro\metre}
\end{equation}

The aperture diameters were then evaluated as offsets above the nominal value in multiples of the clearance. Extensive empirical testing identified that aperture dimensions should be in the range:

\begin{equation}
\text{Aperture} = \SI{1.894}{\milli\metre} + \text{ m } \times \text{Clearance}
\end{equation}

where m is a multiple of our clearance factor. A total of 12 apertures were produced, with three apertures made for each clearance size of 1.894 mm, 1.905 mm, 1.917 mm and 1.928 mm. These apertures exhibited a size variation of $\pm$ \SI{0.1}{\milli\metre}, aligned with the permissible error limit, alongside the standard aperture size. All apertures were machined with entry and exit chamfers to minimise surface-edge interference and aid fibre guidance.\\

Figure~\ref{fig:circlefit} shows the measurement of the smallest circles that fit within the aperture corners, used for quality assessment. These inspections were critical for verifying the dimensional conformity and suitability of the fabricated apertures.

In Figure~\ref{fig:circlefit}, the radii of the corner-fit circles are larger than the fibre radius (\SI{39.5}{\micro\metre}). Our goal is to arrange the fibres in a hexagonally close-packed lattice. In an ideal lattice, circles fit neatly across the width of the aperture. However, the flat walls of the hexagon meet at angles, leaving small triangular voids that do not perfectly accommodate circular packing.\\

If the effective corner radius is larger than the diameter of the fibre, the extra space allows the fibres to shift, disrupting uniform packing. Conversely, if the corner radius is smaller than the diameter of the fibre, the fibres cannot intrude into the voids, ensuring that the lattice remains tight and all fibres are constrained by the aperture boundaries. This prevents wasted space where fibres could rattle or be misplaced, which is important for maximising fill factor and uniformity in optical fibre bundles.\\

The aperture sizes that were cleared for use were the 1.894 mm, 1.931 mm, 1.908 mm, and 1.901 mm; all the other apertures had large discrepancies, either the smoothness of the inside wall of the hexagons being too rough and could potentially bend and break the fibres during the insertion or the corners of the hexagons of the apertures were too large meaning the apertures were not tight enough.

\subsection{Fibre Handling and Insertion}\label{subsec:insertion}

In each instance of attempting to insert fibres into an aperture, precisely 547 fibres of uniform length are cut. According to \cite{Allington-Smith2013, eigenbrot2012impact}, applying stress to the fibres can result in focal ratio degradation (FRD). Hence, the utilization of a cleaver for fibre cutting is essential, as it mitigates stress or possible damage to the fibres and decreases the roughness of the terminated surface.\\

A packing jig was employed to insert the fibers into the aperture, in addition to attempting manual insertion. Large fibres with strong core diameters (such as SMI 200 (\SI{200}{\micro\metre} core diameter \citep{Sabya2024}) can be inserted by collectively pushing the bundle through a precision machined aperture, an approach that fails for weaker fibres (\SI{50}{\micro\metre} core) that tend to buckle.

\begin{figure}[h!]
\centering
\includegraphics[width=0.55\linewidth]{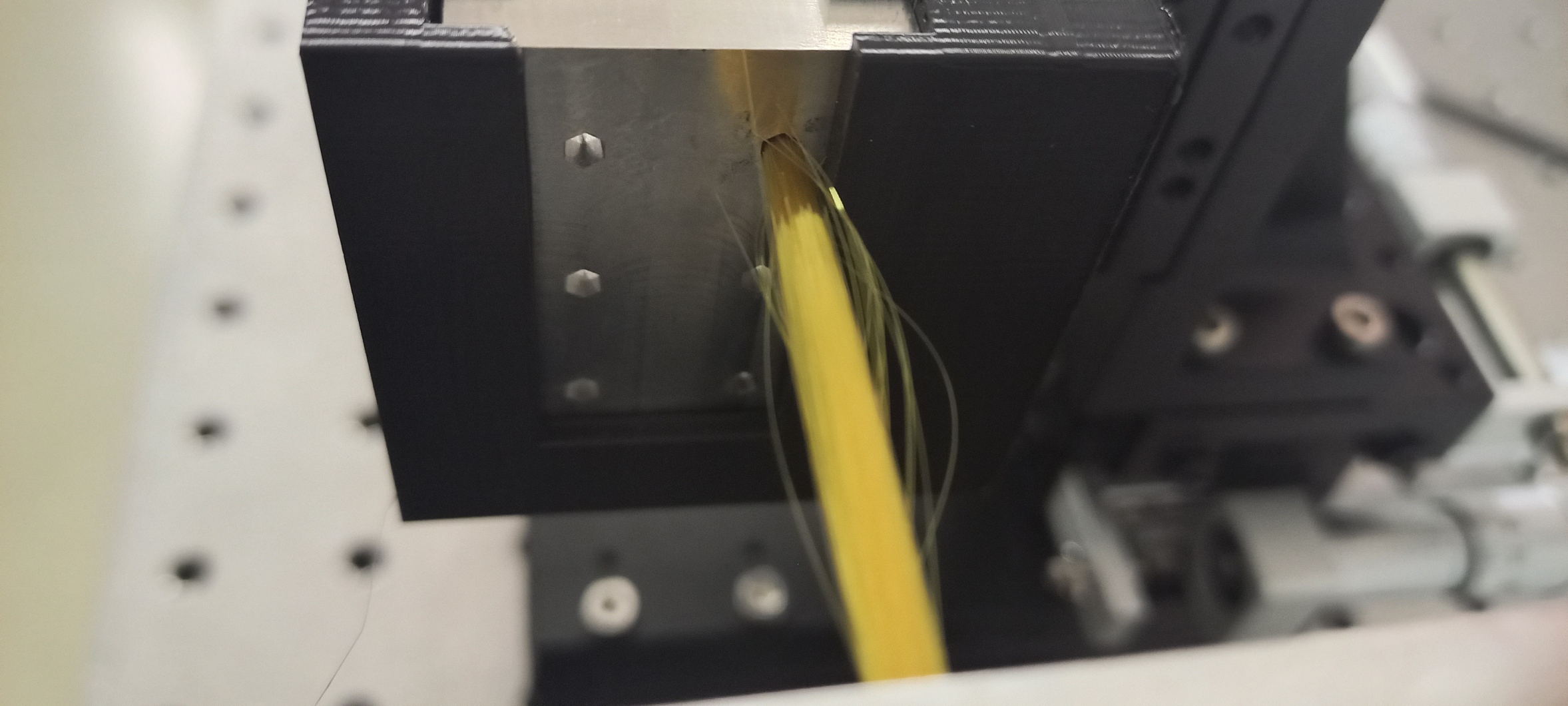}
\caption{Packing jig and aperture holder used for staged fibre insertion.}
\label{fig:packing}
\end{figure}

Initially, the objective was to insert the fibres into the designated aperture with a nominal diameter of 1.894 mm post-removal of the fibre buffer. A solution consisting of 98\% sulfuric acid was employed and heated to a temperature of 180 $^\circ$C. The fibres were agitated to enhance the rate of the chemical reaction. Following the stripping process, the fibres were immersed in deionized water to neutralize the chemical reaction. The buffer-stripped fibres were subsequently consolidated using a heat shrink tube. An attempt was made to insert the stripped section of the fibres through the aperture, followed by the unstripped section. However, the stripped section was prone to breakage, and the aperture was insufficiently sized to accommodate the insertion of the unstripped fibre bundle.\\

Ultimately, isopropyl alcohol (IPA) was utilized to maintain the cohesion of the fibres, thus preventing their dispersion. The fibres were immersed in an IPA solution and placed within an ultrasonic cleaner. The insertion process was facilitated by thermally expanding the aperture slightly. Immersing the fibres in IPA and elevating the temperatures of both the room and aperture allowed for seamless insertion through the 1.931 mm aperture. The aperture size of 1.931 mm was determined to be the minimum feasible configuration for this procedure.

\subsection{Adhesive Application and Curing}\label{subsec:gluing}

\begin{figure}[h!]
\centering
\includegraphics[width=0.3\textwidth, angle = 90]{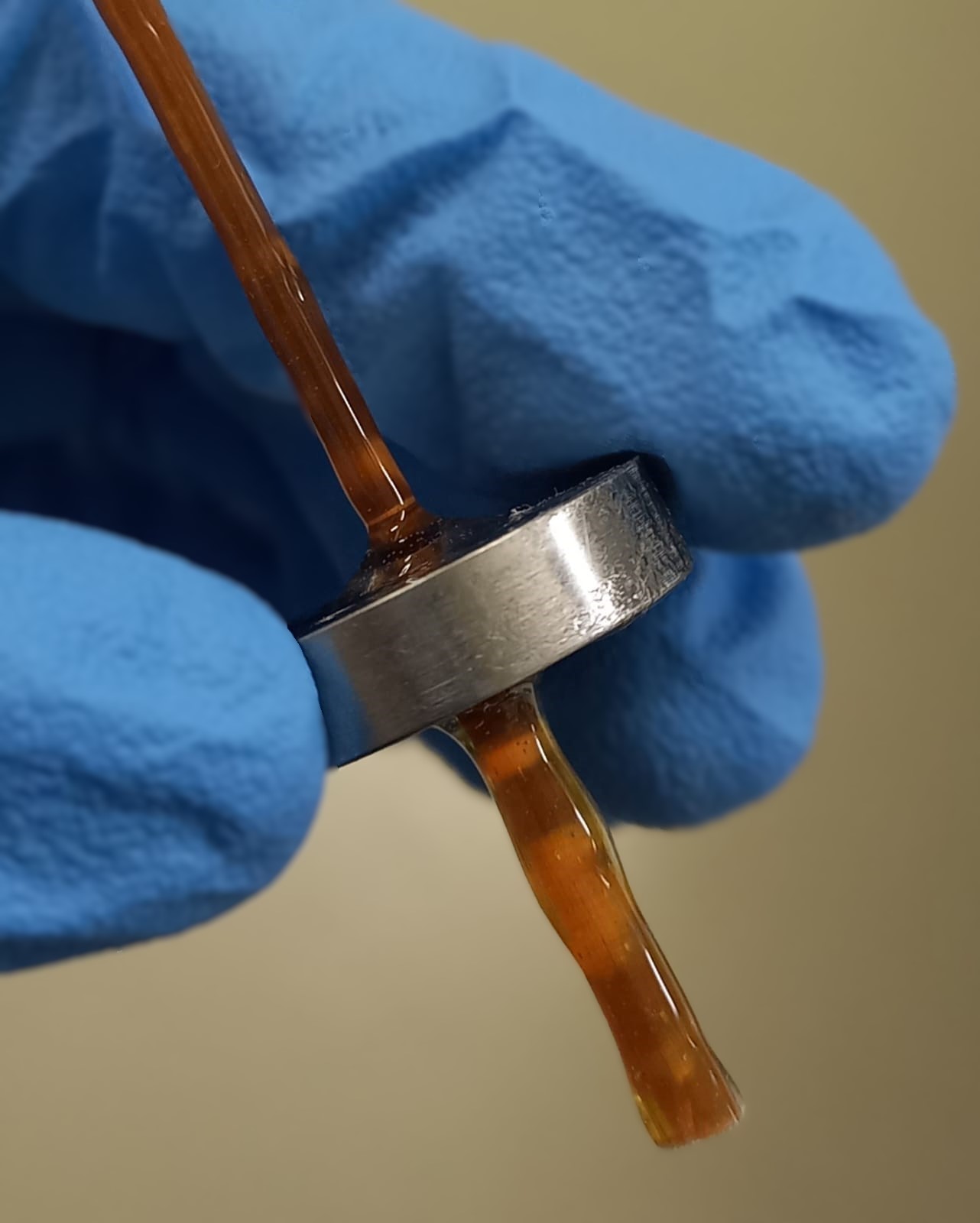}
\caption{Fibre bundle after epoxy curing.}
\label{fig:glued}
\end{figure}

To ensure that the fibres remain intact and placed within, we use a transparent two-part epoxy adhesive 301 2.5GM BI-PAK 80168250001 from \href{https://www.epotek.com/}{Epoxy Technology, Inc (EPO-TEK)}; the epoxy components are mixed within their original packaging prior to opening. After mixing, the two-part epoxy glue is poured into a resealable plastic cup. We carefully used a gauge 18 needle syringe to draw the glue from the cup, minimising the introduction of air bubbles. The glue is applied in phases to the fibre bundle; initially, it is added to the fibres on the aperture's exit side, allowing it to penetrate by gently separating the fibres. The bundle is then slightly retracted through the aperture to facilitate glue flow. The bundle is left undisturbed at room temperature for 48 hours to fully cure. The result is depicted in Figure \ref{fig:glued}.

\subsection{Polishing and Surface Preparation}\label{subsec:polish}

\begin{figure}[h!]
\centering
\begin{subfigure}[t]{0.49\linewidth}
\centering
\includegraphics[width=\linewidth]{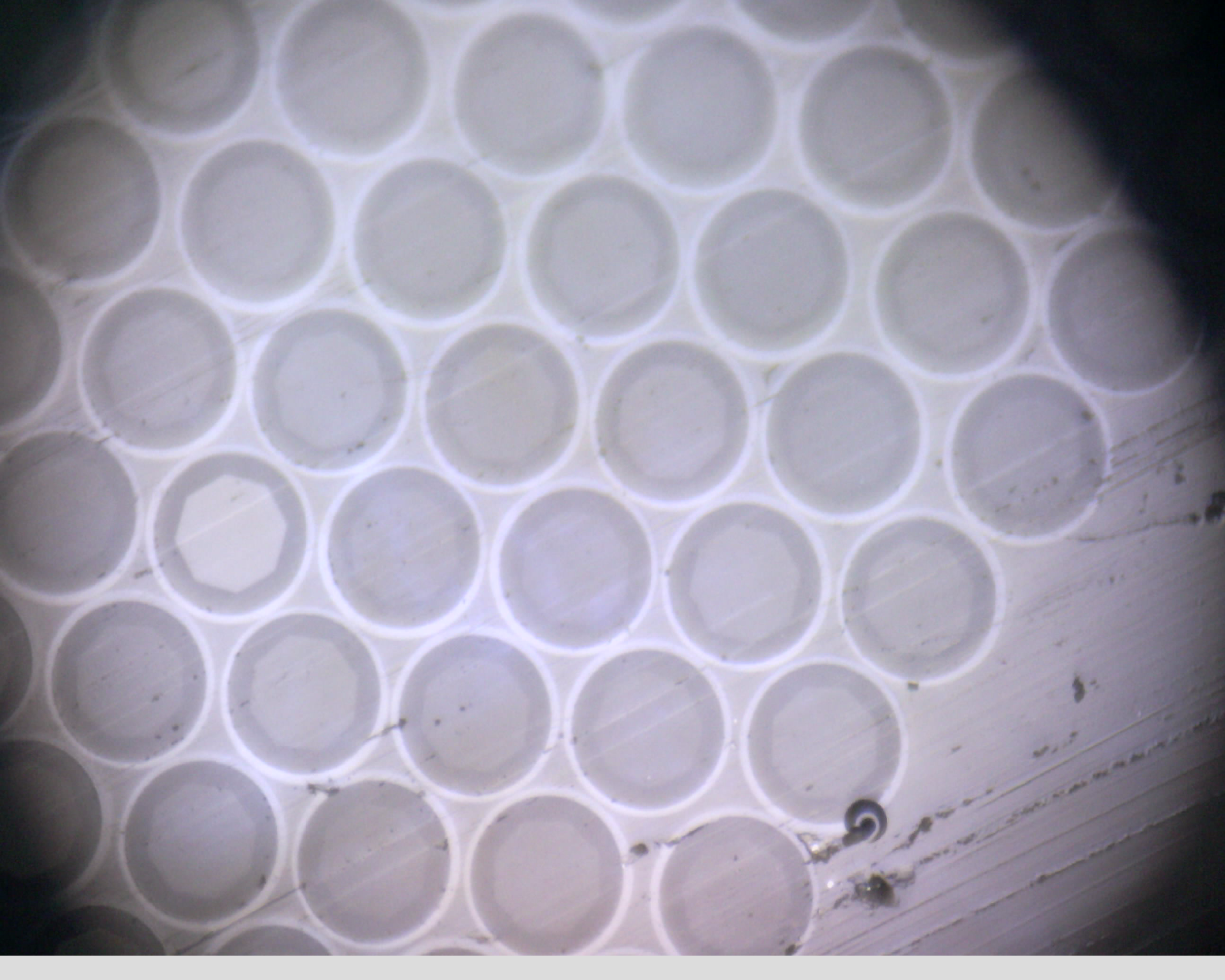}
\caption{Polished fibre bundle end face after the final \SI{1}{\micro\metre} polishing step. The surface is generally smooth, though a small chip is visible on one fibre end, and faint scratches greater than \SI{1}{\micro\metre} in length remain. These features indicate that while the polishing process largely removed irregularities from earlier stages, some minor defects persist.}
\label{fig:polish}
\end{subfigure}
\hfill
\begin{subfigure}[t]{0.48\linewidth}
\centering
\includegraphics[width=\linewidth, height =0.8\linewidth ]{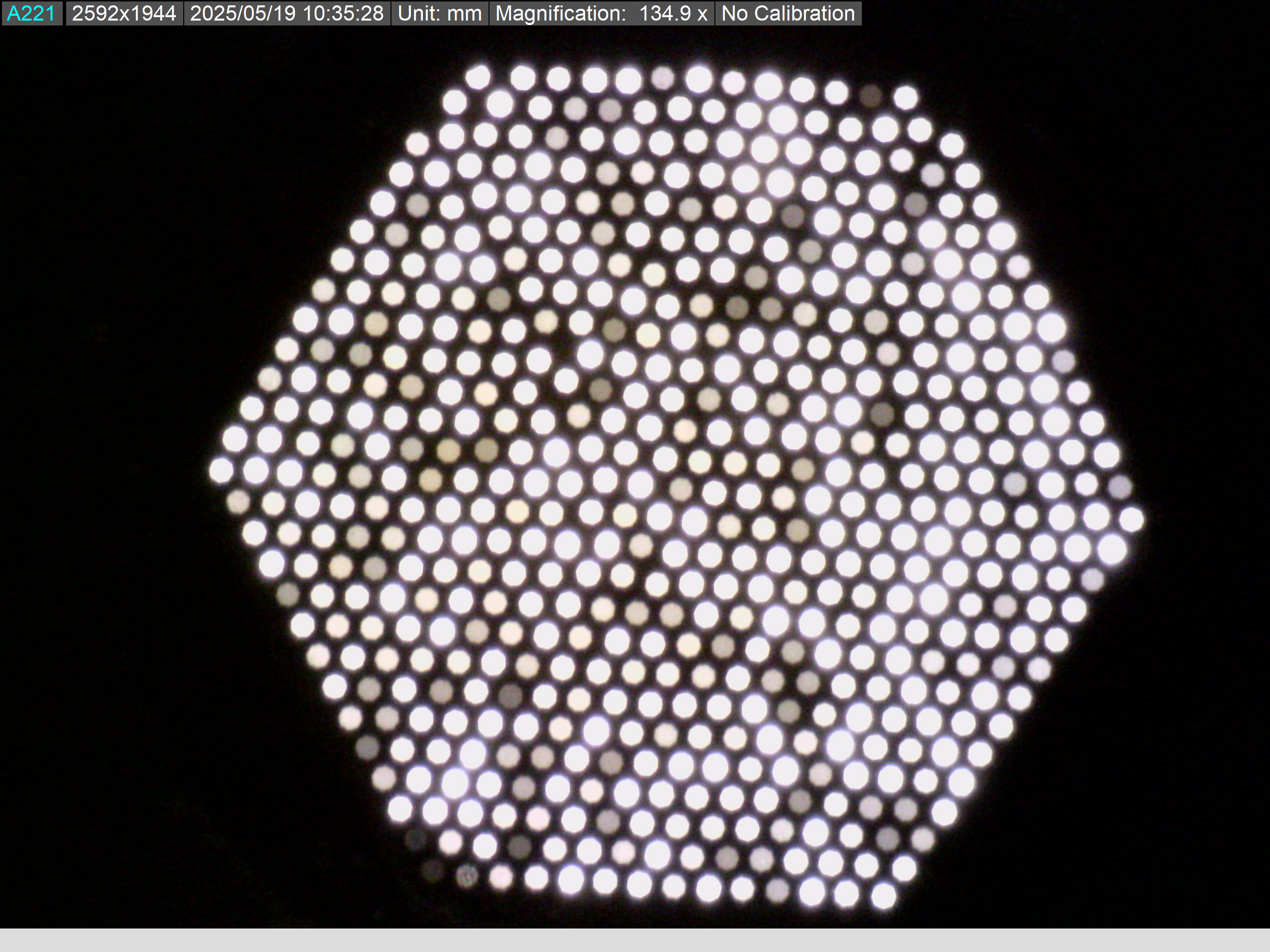}
\caption{Overall view of the completed fibre bundle after the full multistep polishing procedure. Most fibres appear bright, but several are noticeably dimmer, with approximately three fibres showing little or no transmission. Small gaps between fibres and slight positional shifts are also evident, indicating non-ideal packing uniformity.}
\label{fig:bundle}
\end{subfigure}
\caption{Results of the fibre bundle polishing process. (a) Microscopic view of the end face after the final \SI{1}{\micro\metre} polish, showing an overall smooth surface but with minor scratches and a small chip at one fibre end. (b) Macroscopic inspection of the fully polished bundle, where non-uniform brightness, dark fibres, inter-fibre gaps, and slight fibre misalignments are visible. These features highlight the influence of polishing quality and packing efficiency on optical performance.}
\label{fig:sidebyside}
\end{figure}

The fibre bundle was subjected to a thorough multistep polishing process using various micron-scale polishing disks, as described by \cite{Smith2020}. The procedure began with a coarse polishing stage, followed by successive steps using disks with grit sizes of 30, 15, 5, 3, and \SI{1}{\micro\metre}. Each stage lasted 2 hours at respective rotational speeds of 5, 4.5, 4, 3.5, and 1 revolution per second, following the polishing protocol described by \cite{Smith2020}. Adjustments in vertical load, speed, and polishing angle were introduced throughout the process to correct surface imperfections, including scratches and uneven patches. These irregularities were traced to the polishing setup rather than the fibre bundle itself. Iterative application of these refinements resulted in a smooth and uniform surface finish. The effectiveness of the final polishing step can be observed in Figure~\ref{fig:polish}, while the overall condition of the fully polished fibre bundle is shown in Figure~\ref{fig:bundle}.\\

The quality of the polished ends of the fibres plays a crucial role in determining the amount of light that passes through a bundle of fibres and its scattering characteristics. Figure~\ref{fig:polish} shows how a small chip on one fibre and faint scratches larger than \SI{1}{\micro\metre} illustrates the amplification of edge scattering and the reduction in performance due to the remaining surface flaws. On the scale of the entire bundle (Figure~\ref{fig:bundle}), inconsistencies in brightness - including some dim fibres and about three fibres showing minimal or no light transmission - highlight the effects of polishing and packing imperfections. Small gaps and local misalignments between fibres decrease the effective packing density, allowing light to scatter into empty spaces rather than being transmitted efficiently. These structural factors add to the inherent attenuation processes such as absorption, contamination, or microbending, resulting in dimmer fibres within the bundle. Although visual examination can identify qualitative defects such as chips, scratches, voids, and variations in brightness, a quantitative evaluation of optical efficiency still necessitates direct throughput measurements. Apparent brightness alone cannot confirm efficiency, although it provides an early indication of polishing quality and bundle integrity. 

\section{Results}\label{sec:results}

Measurement and analysis in error of the accuracy of the positioning of the fibres along the x and y axes, respectively, were performed as detailed by Gaedie et al. \cite{10.1117/12.3063918}. Fibre centroids were extracted from high-resolution microscope images using a custom Python-based image processing pipeline that applied thresholding, morphological filtering, and ellipse fitting. These detected positions were compared with an ideal hexagonal grid or an adaptively generated irregular grid, allowing for precise quantification of fibre placement deviations.\\

For the AMASE-SAAO fibre bundle, positional error analysis yielded standard deviations of $\rm \sigma_X$ = \SI{14.79}{\micro\metre} and $\rm \sigma_Y$ = \SI{11.10}{\micro\metre}, with a radial error of $\rm \sigma_r$ = \SI{9.50}{\micro\metre}. These deviations reflect the practical limitations of packing core fibres \SI{50}{\micro\metre} with low bending stiffness into a confined geometry, highlighting the importance of precise aperture machining and insertion techniques to minimise misalignment of the fibres.

\section{Discussion}\label{sec:discussion}
This project aimed to produce and evaluate a high-density fibre bundle for the AMASE spectrograph. Despite numerous efforts and extensive design revisions, the project did not meet its anticipated performance criteria, largely due to difficulties with accurately inserting fibres into apertures with precise tolerances. One key issue was inserting fibres into apertures that were either too tight or inadequately fitted. The required extremely fine tolerances further complicated the process. Redesigns involved enlarging aperture hole sizes, experimenting with different fibre types such as serum optical polymicro fibres, and trying varied insertion techniques, including chemical stripping and ultrasonic cleaning. However, achieving successful insertion with sufficient positional accuracy remained elusive. Even when fibres were successfully inserted, as in a 1.931 mm aperture, inconsistencies in the packing, ranging from 20 to \SI{25}{\micro\metre}, greatly exceeded the precision $\pm$ 3 \SI{3}{\micro\metre} required for astronomical use. The uneven internal walls of the apertures further complicated the insertion process, and although polishing somewhat improved the situation, issues such as poor fill factors and uneven bundle faces persisted. A significant finding was related to the handling of fibres with varying diameters. For example, sturdier fibres, such as those in the SMI-200 bundle (\SI{240}{\micro\metre} in diameter with buffer), were easier to insert without bending due to their rigidity. Their structural strength and easier creation of smooth inner walls for larger holes facilitated a straightforward assembly process. In contrast, the smaller diameter fibres used in the AMASE bundles (\SI{79}{\micro\metre}) posed considerable challenges during fabrication and assembly, being more susceptible to bending and damage during insertion.

\section{Conclusion}\label{sec:conclusion}
 The development of a fibre instrumentation cable for AMASE was ultimately unsuccessful, despite improving our understanding of the physical and mechanical limits of the fibre bundle assembly; the project did not meet the required optical and positional criteria. The key conclusions are as follows.\begin{itemize}
  \item Injection of fibres into tight apertures with micrometre-level tolerance proved ineffective because of their mechanical fragility and size inconsistencies.
  \item The packing precision did not meet the standard $\pm$ \SI{3}{\micro\metre}, with deviations often exceeding \SI{20}{\micro\metre}.
  \item The polishing and termination processes added variability, which affected the uniformity of the fibre face.
\end{itemize}These results reveal that conventional manual fabrication methods are not adequate for producing ultradense, small-core fibre bundles with high precision. Future work should investigate the following,\begin{itemize}
  \item Employ laser micro-machining or precision machining techniques for aperture fabrication.
  \item Adopt alternative designs such as individually machined fibre holes to prevent collective misalignment.
  \item Using alignment jigs and automated insertion tools to improve repeatability.
  \item Enhancing adhesive application techniques to ensure consistent distribution and reduce stress.
\end{itemize}

Although the project did not deliver the intended instrument, it has provided essential insight into the current limitations of fibre bundle fabrication methods. These findings will inform the next development phase for AMASE or similar astrophotonic instrumentation projects.


\bibliographystyle{IEEEtran}
\bibliography{iopart-num}

\newpage
\appendix
\section*{APPENDIX A}
\begin{figure}[h!] 
    \centering
    \resizebox{\textwidth}{!}{\includegraphics{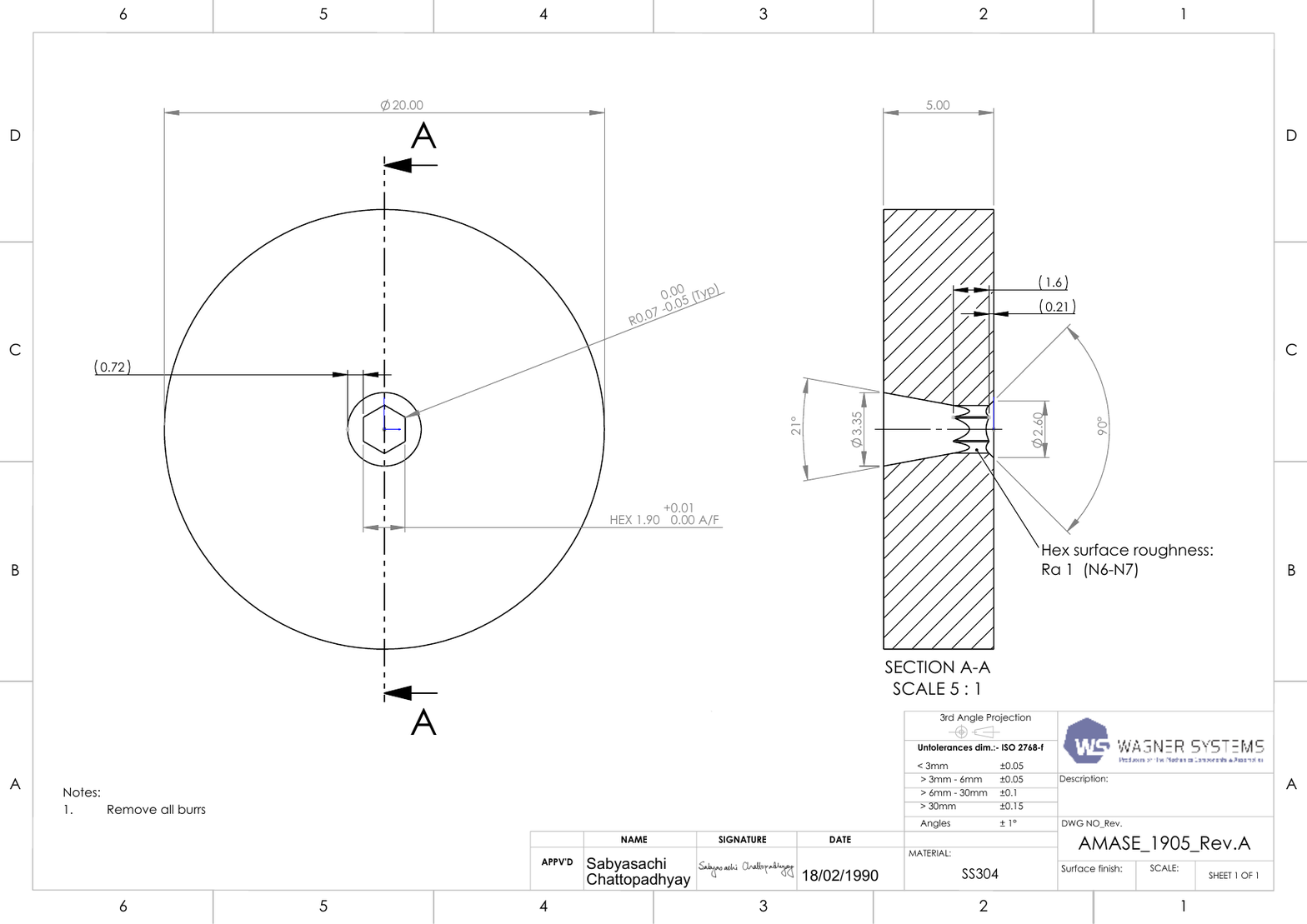}}
    \caption{This CAD illustration showcases the 1.931 mm aperture, emphasizing the design of our stainless steel hexagonal fibre-holding apertures. These apertures are round, with a diameter of 20 mm and a thickness of 5 mm. On the flat circular surface displayed above, a chamfer is seen at the exit of the hexagon hole on the aperture. This chamfer is intentionally crafted to assist the fibres in exiting smoothly when passed through the aperture's opening. To the right, a sectional view (Section A-A) of the aperture shows a significant chamfer on the entrance side (3.35 mm wide), which simplifies the insertion into the aperture hole. In the central section (1.6 mm height), the hexagon is exactly 1.931 mm to hold the fibre securely in place, reducing the likelihood of fibre movement. This section is then followed by an exit side chamfer (0.21 mm height) that is visible on the left side's flat circular surface.}
    \label{AppendixA}
\end{figure}
  
\end{document}